%% file: main.tex
\documentclass[a4paper,onecolumn,11pt,accepted=2025-01-16]{quantumarticle}
\pdfoutput=1
\usepackage[T1]{fontenc}
\DeclareSymbolFont{calletters}{OMS}{cmsy}{m}{n}
\DeclareSymbolFontAlphabet{\mathcal}{calletters}

\usepackage{fullpage}
\usepackage{amsmath}
\usepackage{amsthm}
\usepackage{amsfonts}
\usepackage{amssymb}
\usepackage[dvipsnames]{xcolor}
\usepackage{hyperref}
\usepackage[numbers]{natbib}
\usepackage{svg}
\usepackage{verbatim}
\usepackage{lipsum}
\usepackage{braket}
\usepackage[noabbrev,capitalize]{cleveref}
\usepackage{threeparttable}
\usepackage{qcircuit}
\usepackage{svg}
\usepackage{microtype}
\usepackage{thm-restate}
\usepackage{wrapfig}
\usepackage[normalem]{ulem}
\hypersetup{
  colorlinks = true,
  linkcolor = {MidnightBlue},
  citecolor = {MidnightBlue},
  urlcolor  = {MidnightBlue}
}
\usepackage{listings}\newtheorem{theorem}{Theorem}[section]
\newtheorem{lemma}[theorem]{Lemma}
\newtheorem{definition}[theorem]{Definition}
\newtheorem{corollary}[theorem]{Corollary}

\newtheorem{proposition}[theorem]{Proposition}

\newcommand{\N}{\mathbb{N}}
\newcommand{\calP}{\mathcal{P}}
\newcommand{\calG}{\mathcal{G}}

\newcommand{\calU}{\mathcal{U}}

\newcommand{\calC}{\mathcal{C}}
\newcommand{\calT}{\mathcal{T}}

\newcommand{\calI}{\mathcal{I}}
\newcommand{\stkout}[1]{\ifmmode\text{\sout{\ensuremath{#1}}}\else\sout{#1}\fi}
\newcommand{\edit}[2]{#2}

\newcommand\norm[1]{\lVert#1\rVert}

\newcommand{\ceil}[1]{\left\lceil #1 \right\rceil}
\renewcommand{\epsilon}{\varepsilon}

\newcommand\blfootnote[1]{%
  \begingroup
  \renewcommand\thefootnote{}\footnote{#1}%
  \addtocounter{footnote}{-1}%
  \endgroup
}

\begin{document}

\title{Tight Bounds on the Spooky Pebble Game: Recycling Qubits with Measurements}

\author{Niels Kornerup*}
\affiliation{Department of Computer Science, The University of Texas at Austin, United States of America}
\orcid{0000-0002-1519-726X}
\author{Jonathan Sadun}
\affiliation{}
\author{David Soloveichik*}
\affiliation{Department of Electrical and Computer Engineering, The University of Texas at Austin, United States of America}
\orcid{0000-0002-2585-4120}

\maketitle
\blfootnote{$^*$Research supported by Schmidt Science Polymath award}
\input{abstract}
\input{1-introduction}
\input{2-preliminaries}
\input{3-spooky-pebbling-the-line}
\input{4-lower-bounds}
\input{4.3-tree-lb-proof}
\input{6-DAG-spooky-pebble}
\input{future-work}
\input{7-acknowledgements}
\bibliographystyle{quantum}
\bibliography{sources}

\appendix

\end{document}

%% file: abstract.tex
\begin{abstract}
Pebble games are popular models for analyzing time-space trade-offs.
In particular, reversible pebble game strategies are frequently applied in quantum algorithms like Grover's search 
to efficiently simulate classical computation on inputs in superposition, 
as unitary operations are fundamentally reversible.
However, the reversible pebble game cannot harness the additional computational power granted by intermediate measurements, which are irreversible.
The spooky pebble game, which models interleaved Hadamard basis measurements and adaptive phase corrections, reduces the number of qubits beyond what purely reversible approaches can achieve.
While the spooky pebble game does not reduce the total space (bits plus qubits) complexity of the simulation, it reduces the amount of space that must be stored in qubits.
We prove asymptotically tight trade-offs for the spooky pebble game on a line with any pebble bound. This in turn gives a tight time-qubit tradeoff for simulating arbitrary classical sequential computation when using the spooky pebble game. For example, for all $\epsilon \in (0,1]$, any classical computation requiring time $T$ and space $S$ can be implemented on a quantum computer using only $O(T/ \epsilon)$ gates and $O(T^{\epsilon}S^{1-\epsilon})$ qubits.
This improves on the best known bound for the reversible pebble game with that number of qubits, which uses $O(2^{1/\epsilon} T)$ gates.
For smaller space bounds, we show that the spooky pebble game can simulate arbitrary computation with $O(T^{1+\epsilon} S^{-\epsilon}/\epsilon)$ gates and $O(S / \epsilon)$ qubits whereas any simulation via the reversible pebble game requires $\Omega(S \cdot (1+\log(T/S)))$ qubits.

We also consider the spooky pebble game on more general directed acyclic graphs (DAGs), capturing fine-grained data dependency in computation.
\edit{We show that this game can outperform the reversible pebble game on trees. Additionally any DAG can be pebbled with at most one more pebble than is needed in the irreversible pebble game, implying that finding the minimum number of pebbles necessary to play the spooky pebble game on a DAG with maximum in-degree two is PSPACE-hard to approximate.}
{We show that for an arbitrary DAG even approximating the number of required pebbles in the spooky pebble game is PSPACE-hard.
Despite this, we are able to construct a time-efficient strategy for pebbling binary trees that uses the minimum number of pebbles.}
\end{abstract}

%% file: 1-introduction.tex
\section{Introduction}
Pebble games provide a convenient abstraction for reasoning about space and time usage in computation.
Pebble games were first used in \cite{Set73} to determine optimal register allocation for computing straight line programs.
In \cite{HPV77} the authors applied the irreversible pebble game to show that any computation running on a Turing machine in time $T(n)$ can be executed on another Turing machine with space $T(n)/\log T(n)$.
Since then it has found uses in establishing time-space trade-offs for computing functions including matrix vector products (\cite{Tom80}) and memory hard functions based on hashing (eg. \cite{PTR+76, LT82, DNW05, RD16, BZ17}).
In \cite{Ben89}, Bennett used a reversible pebble game to give a general mechanism for reversibly simulating irreversible computation.
As an example of more recent work,
\cite{BHL22} applied the reversible pebble game to analyze the post-quantum security of these memory hard functions.

In \cite{Gid19} Gidney introduced a new \emph{spooky pebble game} to study time-qubit trade-offs in quantum simulation of classical computation on inputs in superposition.
Many quantum algorithms use simulation of classical computation in superposition as a subroutine.
For example, applying Grover's search \cite{Gro96} to find an $x$ such that $f(x) = 1$, requires a quantum circuit that implements the following mapping:
\begin{equation*}
    \sum_x \alpha_x \ket{x}\ket{j_x} \to \sum_x \alpha_x \ket{x}\ket{j_x \oplus f(x)}
\end{equation*}
which represents the evaluation of $f$ on inputs in superposition.
Since quantum gates are unitary (and therefore reversible) operations, $f(x)$ has traditionally been simulated with a reversible circuit.
This comes with a time-space overhead that is often overlooked.
The spooky pebble game uses intermediate measurements to make such a simulation more efficient than would be possible using reversible circuits.
We expand upon this pebble game and show tight time-space trade-offs for how efficiently it can simulate classical computation.

While the spooky pebble game can reduce the number of qubits needed to perform a computation, it is worth noting that it introduces new classical ancillary bits and does not reduce the total memory (qubits plus classical bits). Nonetheless, qubits are a much more limited resource than classical bits \cite{OCC02}. As such, we believe that this trade-off makes the spooky pebble game pragmatic for designing algorithms that run on quantum computers.

\paragraph{Previous work}

For any $\epsilon \in (0,1]$ the reversible pebble game can be used to reversibly simulate any irreversible computation that runs in time $T$ with $S$ space using $O(T^{1+\epsilon}S^{-\epsilon})$ steps and $O(S (1+\log (T/S)))$ space \cite{Ben89, LS90}.
However, the asymptotic notation above hides a constant factor cost of approximately $\epsilon2^{1/\epsilon}$ in the space term \cite{LS90}.

It was also shown that, using the reversible pebble game, space $O(T^\epsilon S^{1-\epsilon})$ is sufficient to simulate irreversible computation in linear time, but similarly, this result features a steep but constant $2^{1/\epsilon}$ cost in the time of the simulation \cite{Kra01}.
Kr{\'a}l'ovi{\v{c}} also showed that any simulation via reversible pebbling the line with $O(S \cdot (1+\log T/S))$ qubits must use $\Omega(T \cdot (1+ \log T/S))$ steps---a lower bound that is not achieved by any known reversible pebbling algorithm \cite{Kra01}.
In \cite{BHL22} the authors used a parallel version of the reversible pebbling game to show (parallel) time-space efficient algorithms for computing cryptographic objects known as data-independent memory hard functions.

Other works have tried to directly improve the qubit efficiency of running classical subroutines on a quantum computer without going through reversible simulation.
In \cite{PJ06} the authors showed how a classically controlled quantum Turing machine can simulate a classical Turing machine with no loss in time or space complexity.
In \cite{AMP02, CKP13} the authors showed that one qubit is sufficient to simulate $\text{NC}^1$ in polynomial time.

In \cite{Gid19}, Gidney introduced the idea of measurement-based uncomputing with his spooky pebble game.
The spooky pebble game extends the reversible pebble game by allowing intermediate measurements that enable irreversible behavior.
The pebble game is called spooky because these measurements have a chance to produce undesired phases (or ghosts) that need to be removed before completing the computation; otherwise they will obstruct the desired interference patterns generated by subsequent gates.
Gidney showed how the spooky pebble game can be used to simulate any classical computation with only a constant factor blowup in space and a quadratic blowup in the running time, which is impossible for reversible computation \cite{Kra01}. The spooky pebble game was recently extended to work on arbitrary DAGs in \cite{QL23, QL24}, similar to the reversible pebble game.
These works developed a SAT solver that can compute the minimum runtime necessary to solve the spooky pebble game on an arbitrary DAG. In \cite{QL24} the authors recently proved that deciding if a DAG can be pebbled with $s$ pebbles is a PSPACE-complete problem, although their hard case requires a DAG with maximum in-degree $s-1$.

\paragraph{Our results}
We build on the spooky pebble game framework introduced by Gidney (\cite{Gid19}) and prove asymptotically tight upper and lower bounds on spooky pebbling the line.
We do this by proving that---for any pebble bound---there always exists an optimal strategy that has a specific form, constructing such an algorithm, and analyzing a recurrence relation to lower bound the runtime of such algorithms.

We delineate the entire achievable frontier of the spooky pebble game and our tight trade-off bounds can be applied in many different regimes. 
For example, for any $\epsilon \in (0,1]$, any classical computation that runs in $T$ time with $S$ space can be simulated on a quantum computer using only $O(T/ \epsilon)$ steps and $O(T^{1+\epsilon}S^{1-\epsilon})$ qubits.
This is an exponential improvement in $\epsilon$ over the reversible bound in \cite{Kra01}.
We also show that any computation can be simulated in $O(T^{1+\epsilon}S^{-\epsilon}/\epsilon)$ steps with only $O(S / \epsilon)$ qubits, which is fewer qubit than would be possible (independent of the time bound) with reversible pebbling \cite{LV96, LTV98, Kra01}.
Interestingly, when $\epsilon = 1/\log (T/S)$, this matches the (unobtained) lower bound for reversible pebbling proved in \cite{Kra01}.

We then show that any \edit{}{ir}reversible pebbling can be converted to a spooky pebbling using only one additional pebble.
This gives us that, in general, the number of pebbles needed for the spooky pebble game is PSPACE-hard to approximate\edit{, even when only considering graphs with a maximum in-degree of two}{}.
\edit{}{Finally we }discuss the spooky pebble game on directed acyclic graphs (DAGs) where we show that it is possible to spooky pebble the complete binary tree \edit{}{of height $h$} \edit{with $n=2^h -1$ nodes}{(i.e., $n=2^h -1$ nodes)} using $h+1$ pebbles and $O(n \log n)$ steps\edit{,}{.} \edit{which}{This} is less pebbles than is possible in the reversible pebble game \edit{.
This addresses}{addressing} an open question posed in \cite{QL24} regarding efficient algorithms for the spooky pebble game on trees.

\begin{figure*}
\centering
\begin{threeparttable}
\begin{tabular}{rlll}
Source & Game & Pebbles & Steps \\
\hline
& irreversible & $2$ & $O(n)$ \\
\cite{Gid19} & spooky & $3$ & $O(n^2)$ \\
\cref{cor-low-pebbles-pebbling} & spooky & $2 + 1/\epsilon$ & $O(n^{1 + \epsilon} / \epsilon)$ \\
\cite{Ben89} & reversible & $O(\epsilon 2^{1/\epsilon} \log n)$ \cite{LS90} & $O(n^{1 + \epsilon})$ \\
\cite{Gid19}, \cref{cor-log-pebbles-pebbling} & spooky & $O(\log n)$ & $O(n \log n)$ \\
\cite{Kra01} & reversible & $O(n^\epsilon)$ & $O(2^{1/\epsilon} n)$ \\
\cref{cor-lin-time-pebbling} & spooky & $O(n^\epsilon)$\tnote{\textdagger} & $O(n / \epsilon)$ \\
\hline
\cref{lem-spooky-pebble-alg} & spooky & $s$ & $O(mn)$\tnote{\textdaggerdbl}
\end{tabular}
\begin{tablenotes}
\item[\textdagger] When $(2/\epsilon)^{2/\epsilon} \leq n$
\item[\textdaggerdbl] Where $\binom{m + s - 2}{s - 2} \geq n$
\end{tablenotes}
\caption{Summary of results on pebbling the line of length $n$. All asymptotics hold simultaneously for $n \to \infty$ and $\epsilon \to 0$.}\label{tab-line-pebblings}
\end{threeparttable}
\end{figure*}

%% file: 2-preliminaries.tex
\section{Preliminaries}

\paragraph{Reversible Computation}
We say that a computation is \emph{reversible} if, after every time step, there is a unique predecessor state for the computation; a Turing machine that moves from left to right inverting the bits on its tape is reversible while one that sets its tape to zero is not.
When a series of quantum transformations are applied to a quantum system, it results in a unitary transformation $U$ being performed on the system. So long as no measurements are taken, the original state of the quantum system can be restored from the final state by applying the inverse $U^\dagger$ \cite{RP11}. Thus measurement-free quantum computation must be reversible. 
This means that implementing classical algorithms using a quantum computer often implicitly involves the additional step of making the algorithm reversible.

Reversible computation is also important when considering energy-efficient classical computation.
Thermodynamics gives computation an energy lower bound of $kT \ln 2$ per irreversible step---a barrier that can be circumvented by making computation reversible \cite{Lan61,Ben73}.
Unfortunately, all known general techniques for converting irreversible algorithms to reversible ones feature an asymptotic time or space overhead \cite{Ben89, LMcT00, SM13, AGS15}. In fact, relative to random oracles there is a provable separation between time-space trade-offs for reversible and irreversible computation \cite{FA17}.

Of the strategies for making classical computation reversible, the most widely used method is the reversible pebble game \cite{Ben89}. In the reversible pebble game, steps of an irreversible algorithm are simulated reversibly by placing and removing pebbles on a line, or more generally, any directed acyclic graph (DAG). The largest number of pebbles placed at any time
corresponds to the space needed for the simulation and the number of steps corresponds to the time of the simulation.

\paragraph{Uncomputation}
Recycling space is key to space efficiency. 
But while irreversible algorithms can simply erase values whenever they are no longer needed,
such values require more care \edit{of}{to} dispose \edit{}{of} in reversible and quantum computation.
In order to erase a value in a reversible manner, it is necessary to end up in a state where it would be possible to efficiently recompute the deleted value.
Thus \emph{uncomputation}, the reversible analog of deletion, requires access to the same information needed to compute the deleted value.
The importance of uncomputation in quantum circuits it twofold: (1) qubits are an expensive resource for quantum algorithms so it is important to design them so that they can run on as few qubits as possible and (2) failing to uncompute values in a quantum circuit results in entangled garbage that prevents desired interference patterns \cite{PBSV21}.

We say a quantum circuit \emph{uncomputes} a function $f$ if it can perform the following mapping on any quantum state where the $x_i$ are all distinct:
\begin{equation*}
    \sum_i \alpha_i \ket{x_i}\ket{f(x_i)} \to \sum_i \alpha_i \ket{x_i}\ket{0}
\end{equation*}
Note that the standard unitary $U_f$ for computing $f$, which maps $U_f \ket{x}\ket{b} = \ket{x}\ket{b \oplus f(x)}$ is also a unitary that uncomputes $f$.
Carefully balancing computation and uncomputation is vital for designing space and time efficient quantum algorithms.

\paragraph{Ghosting}

Gidney points out that full uncomputation is not always necessary for recycling qubits when simulating classical algorithms in superposition.
Instead of full uncomputation, he developed a clever scheme using intermediate measurements to ``compress'' qubits into classical bits \cite{Gid19}.
We say that a quantum circuit $\calC$ \emph{ghosts} a register if it can perform the following mapping on any quantum state where the $x_i$ are all distinct:
\begin{equation*}
    \sum_i \alpha_i \ket{x_i}\ket{y_i} \xrightarrow{\calC} \sum_i \alpha_i (-1)^{b \cdot y_i} \ket{x_i}\ket{0}
\end{equation*}
and $b$ is a classical bitstring returned by the circuit.
In \cite{Gid19} Gidney calls this $(-1)^{b \cdot y_i}$ phase a \emph{ghost} of $y$, as the logical value of this register has been erased and replaced with a phase. Since the logical value of the register has been zeroed out, it can be used as if it were a fresh ancilla register for any subsequent computation acting only in the standard basis.
However for a quantum circuit to behave correctly, this ghost must be removed before acting on this register in another basis.
By recreating the state of the $y$ register as it was before ghosting and recalling the classical bitstring $b$, the ghost can be removed by applying a $b$-controlled Z gate to the register.
Finally uncomputation can be used to properly uncompute the register.
While this might seem like uncomputation with extra steps, we will see that \emph{reusing registers before we could uncompute their values allows for more qubit and gate efficient quantum algorithms.}

\begin{figure*}
\[
\begin{array}{c}
\Qcircuit {
 & & \cwx[2] & \control \cwx[2] \cw & \cw &\ustick{b\in\{0,1\}} \cw & \cw & \cw &  \control \cw \cwx[2]  & \\
 &  &  &  &  &  & \lstick{\ket{x}} & \multigate{1}{U_f} & \qw & \multigate{1}{U_f} & \rstick{\ket{x}} \qw \\
\lstick{\ket{f(x)}} & \gate{H} & \meter \cwx[-2] & \targ & \qw & \raisebox{3.5em}{$(-1)^{bf(x)}\ket{x}\ket{0}$} \qw & \qw & \ghost{U_f}  & \gate{Z}  & \ghost{U_f} & \rstick{\ket{0}} \qw  \gategroup{1}{2}{3}{4}{1em}{--} \gategroup{1}{8}{3}{9}{1em}{--}
}
\end{array}
\]
\caption{Circuit to ghost $f(x)$ and later correct the phase in order to overall uncompute $f(x)$ using measurement-based uncomputing, as described in \cite{Gid19}.
}
\label{fig-uncompute}
\end{figure*}

Ghosting is not a unitary operation and thus requires intermediate measurements: 
for example, if you tried to ghost the second qubit in the state $(\ket{00}+\ket{01})/\sqrt{2}$ and receive $b=0$, then you would end up with the non-normalized state $(\ket{00} + \ket{00})/\sqrt{2} = \sqrt{2} \ket{00}$.
Importantly, note that ghosting is only defined to perform this mapping when the $x_i$ are distinct, as that is sufficient to prevent interference between basis states and make it a norm-preserving operation.

In \cite{Gid19} Gidney shows that ghosting can be performed by measuring in the Hadamard basis, recording the result as classical output $b$, and then using $b$ to apply a controlled $X$-gate to the register. \cref{fig-uncompute} shows what this circuit looks like, as the gates on the left ``ghost'' the $f(x)$ register and the gates on the right recompute $f(x)$ before removing the ghost and uncomputing $f(x)$.
Importantly, the gates that ghost the $f(x)$ register do not use the $x$ register, so these operations can be performed even when the $x$ register is holding a different value.
However, removing the ghost then requires the presence of $x$.

\begin{figure*}
\centering
\[
\begin{array}{c}
\Qcircuit {
& \cwx[2] &\cw &\ustick{b\in\{0,1\}} \cw &\cw & \control \cw \cwx[1] &\\
 &  &  &  & \lstick{\ket{x}} & \multigate{1}{E_f} & \multigate{1}{U_f} & \rstick{\ket{x}} \qw\\
\lstick{\ket{f(x)}} & \gate{G} & \qw & \raisebox{3.5em}{$(-1)^{bf(x)}\ket{x}\ket{0}$} \qw & \qw & \ghost{E_f} & \ghost{U_f} & \rstick{\ket{0}} \qw
}
\end{array}
\]
\caption{The circuit from \cref{fig-uncompute} using G and $E_f$ circuit macros. \edit{}{Note that the $\bullet$ on the classical wire above the $E_f$ gate represents a controlled application of the component $Z$ gate rather than a controlled application of an entire module.}}
\label{fig-uncompute-G-E-notation}
\end{figure*}

For compactness, we compress the highlighted gates on the left and the right of \cref{fig-uncompute} into single gates $G$ and $E_f$ as shown in \cref{fig-uncompute-G-E-notation}.
When performing a classical subroutine on a quantum computer, the $G$ gate can be applied to any register at any time to reset the logical value of its qubits (ghosting).
However, whenever this is done, the $E_f$ gate must be reapplied later in the circuit when it is possible to recompute the original value in the erased register and remove the phase (unghosting).

\begin{figure*}
\centering
\includegraphics[width=\textwidth]{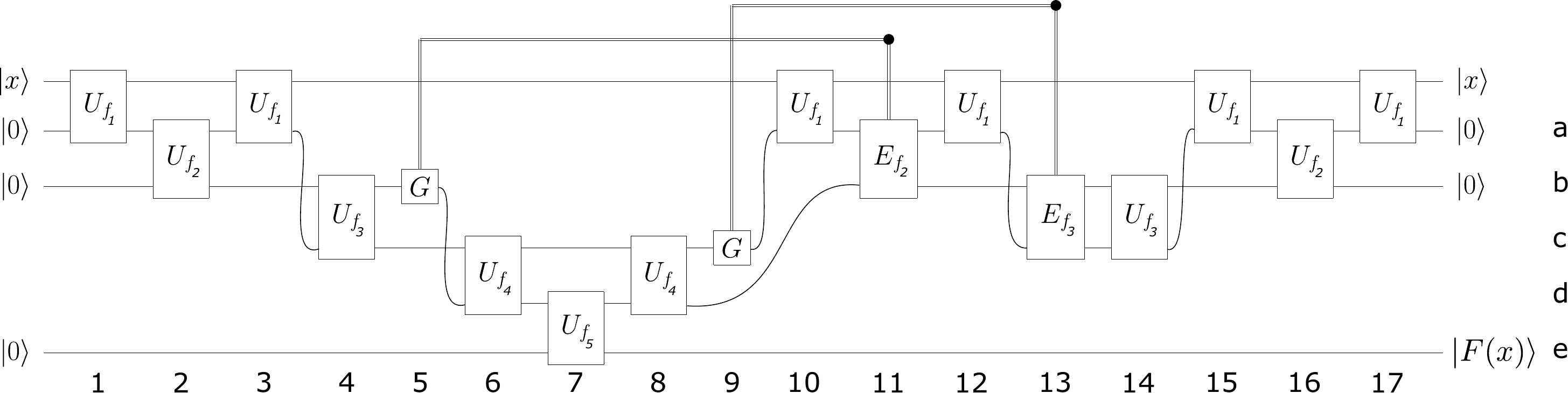}
\caption{Qubit efficient computation of $F(x) = f_5 \circ f_4 \circ f_3 \circ f_2 \circ f_1 (x)$ using ghosting. \edit{}{Again the $\bullet$ on the classical wire above the $E_{f_i}$ gates represents a controlled application of the component $Z$ gate rather than a controlled application of an entire module.}
}
\label{fig-G-E-circuit}
\end{figure*}

\Cref{fig-G-E-circuit} shows how these $G$ and $E_f$ gates can be applied to compute a function with less qubits than would be possible using reversible simulation.
This application is general: we can think of a function $f$ as mapping a computer from its current logical configuration to its configuration at the next time step.
Since the measured value $b$ must be stored from the time a value is ghosted until the time when its ghost is removed, we are not able to reduce the overall space complexity of the simulation;
we are only able to reduce the number of required qubits.

Gidney's original formulation of the spooky pebble game only produced a ghost when the returned value $b \not = 0$ since when $b=0$ no phase is added \cite{Gid19}.
However, we will apply the $G$ gate on groups (registers) of qubits and the probability that all measured qubits yield zero will be negligibly small.
As such, we say that a ghost is always created when we apply the $G$ gate.

\paragraph{Pebble games}
The concept of interweaving computation, uncomputation, and ghosting to efficiently simulate classical computation can be perfectly encapsulated by an appropriate pebble game.
In a pebble game we are given a DAG where the nodes represent the variables involved in a computation and the edges show the dependencies for computing these variables.
For example the computation $c = a + b$ can be represented with three nodes for $a,b,c$ and edges $a \to c$ and $b \to c$.
A pebble can be placed on a node to indicate that the variable is currently stored in a register for the computation.
Thus a pebble can be placed on node $c$ only when there are already pebbles on $a$ and $b$.
Additional restrictions on when pebbles can be placed or removed create different pebble games that can be used to simulate irreversible, reversible, or quantum computation with ghosting.

Such graph representations naturally characterize computations such as straight line programs, circuits, and data-independent cryptographic functions acting on words (registers) of size $w$.
In this setting, pebble games directly model time-space trade-offs---as a pebbling of the DAG for a function $f$ with $s$ pebbles placed concurrently and $\tau$ steps corresponds to a way to compute $f$ with $O(ws)$ bits in $O(\tau)$ steps.

While pebble games are defined for general DAGs, it is possible to think of each pebble as representing a snapshot of the entire state of an arbitrary sequential computation and then represent the computation as a pebble game on a line.
If you can construct a pebbling $\calP_n$ that works on the line of length $n$ in $\tau_n$ steps and uses at most $s_n$ concurrent pebbles,
this leads to a naive way to simulate a computation that requires $T$ time and $S$ space in $O(S \cdot \tau_T)$ steps\footnote{There is a multiplicative term of $S$ here because each state must be copied before simulating the next step.} and $O(S \cdot s_T)$ space (or qubits).
A more clever way to apply pebbling to simulation involves assigning $S$ states to each pebble, which lets us pebble a shorter line without increasing the asymptotic space needed per pebble or the time needed per step.
\begin{proposition}[Implicit in \cite{Ben89}] \label{prop-pebble-to-sim}
Let $M$ be an irreversible sequential machine that computes a function $f$ with $T$ steps using $S$ bits. 
Let $\calP_{T/S}$ be a (spooky) pebbling strategy for the line of length $T/S$ that runs in $\tau_{T/S}$ steps and uses at most $s_{T/S}$ concurrent pebbles.
Then there exists a quantum circuit that can compute $f$ with $O(S\cdot \tau_{T/S})$ gates and $O(S \cdot s_{T/S})$ qubits.
\end{proposition}

We now more formally define the irreversible, reversible, and spooky pebble games that we will use throughout this paper.
Here the $\triangle$ operator denotes the symmetric difference.

\begin{definition}\label{def-irrev-pebble-game}
The irreversible pebble game is a one player game on a DAG $G=(V,E)$ where the goal is to place a pebble on exactly the nodes $T \subseteq V$ \edit{}{called \emph{targets}} with out-degree zero. A pebbling (strategy) is a list of subsets of $V$. $\calP = [\calP_0, \ldots, \calP_\tau]$ where $\calP_0 = \emptyset$ and $\calP_\tau = T$. A strategy is valid as long as
\begin{itemize}
    \item $\norm{\calP_i \triangle \calP_{i+1}} = 1$, and
    \item If $v \in \calP_{i+1} \setminus \calP_i$, then $\text{parents}(v) \subseteq \calP_i$.
\end{itemize}
\end{definition}
We can analyze a pebbling by looking at the number of pebbles and steps it requires.
\begin{definition}
The number of steps $T(\calP)$ in a pebbling strategy $\calP = [\calP_0, \ldots, \calP_\tau]$ is $\tau$.
\end{definition}
\begin{definition}
The number of pebbles $S(\calP)$ in a pebbling strategy $\calP = [\calP_0, \ldots, \calP_\tau]$ is $\max_{i \in [\tau]} \norm{\calP_i}$.
\end{definition}
When restricted to reversible computation, we can modify \cref{def-irrev-pebble-game} to get a pebble game that naturally models the restrictions imposed by reversibility.
\begin{definition}
The reversible pebble game has the same setup as the irreversible pebble game in \cref{def-irrev-pebble-game}. A strategy is valid as long as
\begin{itemize}
    \item $\norm{\calP_i \triangle \calP_{i+1}} = 1$,
    \item If $v \in \calP_{i+1} \setminus \calP_i$, then $\text{parents}(v) \subseteq \calP_i$, and
    \item If $v \in \calP_i \setminus \calP_{i+1}$, then $\text{parents}(v) \subseteq \calP_i$.
\end{itemize}
\end{definition}
The pebble game capturing ghosting admits better pebbling strategies than are possible in the reversible pebble game, assuming we are concerned with the number of qubits rather than the total space complexity.
Given an irreversible pebbling, we can exactly characterize when it produces a ghost.
\begin{definition}\label{def-ghosting-seq}
The ghosting sequence $\calG(\calP) = [\calG_0, \ldots, \calG_{T(\calP)}]$ is defined recursively by $\calG_0 = \emptyset$ and $\calG_{i+1} = (\calG_i \cup \{v \in \calP_i | \text{parents}(v) \not\subseteq \calP_i \}) \setminus \calP_{i+1}$.
\end{definition}
The ghosting sequence cumulatively tracks the locations where pebbles were removed without access to their parents. These are exactly the steps that prevent a strategy from being a valid strategy for the reversible game; a valid reversible pebble game strategy is exactly a strategy with a ghosting sequence consisting only of empty sets. The spooky pebble game relaxes this condition to only requiring the \emph{last} ghost set to be empty. Note that ghosts can be removed by placing a pebble over them.
\begin{definition}
The spooky pebble game has the same setup as the irreversible pebble game in \cref{def-irrev-pebble-game}. A strategy is valid as long as
\begin{itemize}
    \item $\norm{\calP_i \triangle \calP_{i+1}} = 1$,
    \item If $v \in \calP_{i+1} \setminus \calP_i$, then $\text{parents}(v) \subseteq \calP_i$, and
    \item In the ghosting sequence $\calG(\calP)$, the final ghost set $\calG_\tau = \emptyset$.
\end{itemize}
\end{definition}

\begin{wrapfigure}{r}{0.25\textwidth}
\centering
\includegraphics[width=1in]{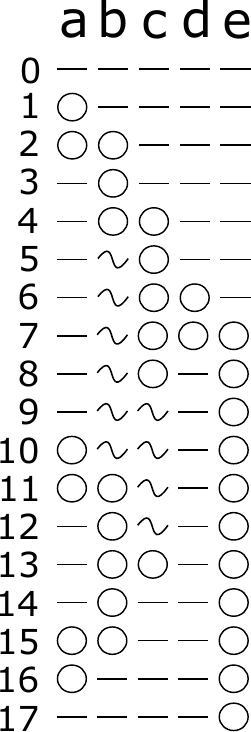}
\caption{A spooky pebbling of the line. Here $\circ$ indicates a pebble and $\sim$ indicates a ghost.
}
\label{fig-quad-pebbling-example}
\end{wrapfigure}

Note that by construction, the spooky pebble game always lies somewhere between the reversible and irreversible pebble games. 
Therefore it is natural to compare it to these other pebble game and investigate when its behavior is closer to that of reversible or irreversible pebbling.

\cref{fig-quad-pebbling-example} gives an example of a spooky pebbling on the line of length $5$ using only $3$ pebbles. Note that this models the same computation performed in \cref{fig-G-E-circuit}.
Each numbered step in \cref{fig-quad-pebbling-example} corresponds to the state of the quantum circuit in \cref{fig-G-E-circuit} after the corresponding numbered gate is applied\edit{.}{; a wire on row c in \cref{fig-G-E-circuit} occurs exactly at the same numbered steps as a pebble at column c in \cref{fig-quad-pebbling-example} and corresponds to storing the value $f_3 \circ f_2 \circ f_1 (x)$ in quantum memory within both models.}
In the reversible pebbling game, this task requires $4$ pebbles. 
The ability to ghost pebbles and remove the ghosts at later steps lets us use one fewer pebble, showing that the spooky pebble game can be used to save qubits.

Rather than describe a pebbling as a list of pebbled nodes, we will sometimes describe algorithms that generate these lists. These algorithms feature the functions ``place'' and ``remove''. The $t$'th call to these functions defines the value of $\calP_{t}$ from $\calP_{t-1}$ as follows:
\begin{itemize}
    \item place($v_i$): $(\calP_{t}) = (\calP_{t-1} \cup \{v_i\})$
    \item remove($v_i$): $(\calP_{t}) = (\calP_{t-1} \setminus \{v_i\})$
\end{itemize}
Intuitively, the place instruction creates a pebble on a node while the remove instruction destroys that pebble. For the spooky pebble game, ghosts are implicitly created by remove instructions according to \cref{def-ghosting-seq}.

%% file: 3-spooky-pebbling-the-line.tex
\section{Spooky pebbling the line}\label{sec-line-pebble}
Before constructing spooky pebbling algorithms, we first define two useful related pebbling tasks.
\begin{definition}
An \emph{unpebbling} is a sequence of place and remove instructions that, assuming the graph starts with pebbles exactly on all nodes with out-degree zero (and no ghosts), results in the graph having neither pebbles nor ghosts on any of its \edit{notes}{nodes}.

An \emph{unghosting} is a sequence of place and remove instructions that, assuming the graph starts with ghosts exactly on all \edit{nodes with out-degree zero (and no pebbles)}{target nodes (nodes with out-degree zero) and no pebbles}, results in the graph having neither pebbles nor ghosts on any of its nodes.
\end{definition}

\edit{}{While an unghosting assumes that the graph only starts with ghosts on nodes with out-degree zero, these sequence of place and remove instructions will remove any additional ghosts that start elsewhere on the DAG, as pebbles must be placed on all nodes in the graph to remove ghosts on the out-degree zero nodes.}
In the rest of this section we restrict our attention to the line graph $G = (\{v_1, \ldots, v_n\}, \{(v_1, v_2), \ldots, (v_{n-1}, v_n)\})$. 
We will canonically refer to $\calP$ as a pebbling algorithm, $\calU$ as an unpebbling subroutine, and $\calU_g$ as an unghosting subroutine.

The spooky pebbling algorithms we discuss in this paper all have a similar recursive form. They start by using a sequence of alternating place and remove instructions to place a pebble on some vertex $v_k$, leave a pebble on this vertex to recursively pebble the rest of the line, and use an unpebbling subroutine to remove the pebble placed at $v_k$. In \cref{sec-lower-bounds} we prove the existence of time-optimal pebbling strategies that employ this structure. The algorithms we discuss in this section are most naturally expressed in the language of unghosting subroutines (where $v_n$ starts as a ghost). These subroutines \edit{and}{} can be converted into pebbling algorithms using the following lemma:
\begin{restatable}{lemma}{unghostingtopebbling}\label{lem-unghosting-to-pebbling}
Let $\calU_g$ be an unghosting algorithm for the line of length $n$ that uses $s$ pebbles and takes $\tau$ steps. Then there exists a pebbling algorithm on the line of length $n+1$ that uses $(s+1)$ pebbles and takes at most $(\tau+1)$ steps.
\end{restatable}
\begin{proof}
We take $\calU_g$ and modify it by adding a pebble instruction on $v_{n+1}$ immediately after the first instruction acting on $v_n$---which must be a pebble instruction. Note that since $v_n$ starts with a ghost, $\calU_g$ must place a pebble at $v_n$ during some step. The resulting algorithm is a valid pebbling that uses at most $\tau+1$ steps and $s+1$ pebbles.
\end{proof}
We now show a simple spooky pebbling algorithm that uses a constant \edit{}{number of} pebbles to pebble the line.
\begin{proposition}[from \cite{Gid19}]\label{lem-quad-alg}
There exists a spooky unghosting algorithm that can \edit{pebble}{unghost} the line graph of length $n$ using $O(n^2)$ steps and only two pebbles.
\end{proposition}
\begin{proof}
Consider the following algorithm:
\begin{lstlisting}[numbers=none]
$\calU_g^{n^2}(\{v_1, \ldots, v_{n}\})$:
    place($v_{1}$)
    if($n=1$):
        remove($v_{1}$)
    else:
        for $i \in [2,n-1]$:
            place($v_{i}$)
            remove($v_{i-1}$)
        place($v_{n}$)
        remove($v_{n}$)
        remove($v_{n-1}$)
        run $\calU_g^{n^2}(\{v_1, \ldots v_{n-1}\})$
\end{lstlisting}
This algorithm removes the ghost on $v_{n}$, leaving a ghost on $v_{n-1}$. This process is repeated until the remaining ghost is on $v_1$, at which point we can remove the pebble without creating a ghost. The number of steps in this pebbling algorithm is given by $T(1) = 2$ and $T(k) = 2k + 2 + T(k-1)$. Unrolling this recursion yields $T(n) = n^2 + 3n - 2$ which is $O(n^2)$.
\end{proof}
We can convert $\calU$ into a pebbling algorithm for the line of length $n$ with $3$ pebbles using \cref{lem-unghosting-to-pebbling}.
\begin{corollary}[from \cite{Gid19}]\label{cor-quad-alg}
There exists a spooky pebbling algorithm that can pebble the line graph of length $n$ using $O(n^2)$ steps and only three pebbles.
\end{corollary}

With the use of additional pebbles, we can create asymptotic improvements in the number of steps used by pebbling algorithms. One additional pebble produces a pebbling that requires only $O(n^{3/2})$ steps. This extra pebble lets us leave a ``checkpoint'' in our computation that makes removing ghosts possible with fewer steps. However this checkpoint must then be removed in an unpebbling subroutine. This idea can be extended to create a general tradeoff between pebbles and steps.
\begin{lemma} \label{lem-spooky-pebble-alg}
For any $s,m \in \N$ such that $n \leq \binom{m+s-1}{s-1}$, there exists an unghosting algorithm for the line of length $n$ that uses at most $s$ pebbles and $O(mn)$ steps.
\end{lemma}

\begin{proof}
We consider the following unghosting algorithm that uses $s$ pebbles for when the line has length exactly $n = \binom{m + s -1}{s-1}$ for some value of $m$:
\begin{lstlisting}[numbers=none]
$\calU_g(\{v_1, \ldots, v_n\},s)$: //Assume that $n = \binom{m + s -1}{s-1}$ for some value of $m$.
    if $(n \leq s)$:
        for $i \in [1,n]$:
            place($v_i$)
        for $i \in [0,n-1]$:
            remove($v_{n-i}$)
    else:
        let $k = \binom{m + s -2}{s-1}$
        place($v_1$)
        for $i \in [2,k]$:
            place($v_{i}$)
            remove($v_{i-1}$)
        run $\calU_g (\{v_{k+1}, \ldots v_n\}, s-1)$ // Line of length $n-k = \binom{m + s-2}{s-2}$.
        remove($v_k$)
        run $\calU_g (\{v_1, \ldots, v_k\}, s)$ // Line of length $k = \binom{m + s -2}{s-1}$.
\end{lstlisting}
Note that if $m'=m-1$ then $k = \binom{m'+s-1}{s-1}$, so our assumption on $n$ holds in the recursive cases.
\edit{}{While the recursive calls to $\calU_g$ may run on lines that contain ghosts on nodes other than the target, removing the ghost at the target will require placing pebbles on all other nodes.
This will remove any additional ghosts present at the start of the recursive call.}
To upper bound the number of steps required to run $\calU_g$ on a line of length $n = \binom{m + s -1}{s-1}$ with $s$ pebbles, we will upper bound the number of steps that act on any node and multiply that by the total number of nodes. Let $T_i(m,s)$ be the number of times $\calU_g$ acts on node $v_i$ and $T^*(m,s) = \max_i T_i(m,s)$. Then directly from the construction of $\calU_g$ we have:
$$T^*(m,s) \leq \begin{cases} 2 & s \geq \binom{m + s -1}{s-1} \\ \max \left(2 + T^*(m-1,s), T^*(m,s-1)\right) & s < \binom{m + s -1}{s-1}.
\end{cases}$$
In other words, when $s \geq n$, each node has a pebble placed and removed at most once.
Otherwise, when $s < n$, the node visited the most times either is before $v_{k+1}$ and gets a pebble placed and removed and is part of the call $\calU_g (\{v_1, \ldots, v_k\}, s)$ or is after $k$ and is part of the call $\calU_g (\{v_{k+1}, \ldots v_n\}, s-1)$.

When $m$ or $s$ is $1$, we observe that $s \geq \binom{m + s -1}{s-1}$, and so the base case $T^*(1,s) = T^*(m,1) = 2$. 
Since every recursive step at most increases $T^*$ by $2$ every time that $m$ increases by $1$, 
it follows that $T^*(m,s) \leq 2m$. This lets us conclude that when the line has length $n=\binom{m+s-1}{s-1}$, we can unghost it using $O(mn)$ steps.

When $n < \binom{m + s -1}{s-1}$ we observe that the above unghosting algorithm can be truncated to the shorter line of length $n$ by ignoring steps on nodes after $v_n$. Again the largest number of steps on any node is at most $2m$ so we can conclude that the unpebbling algorithm takes at most $2mn$ steps.
\end{proof}
We can convert this to a pebbling algorithm with \cref{lem-unghosting-to-pebbling}.
\begin{theorem}\label{thm-spooky-pebble-alg}
    For any $s,m \in \N$ such that $n \leq \binom{m+s-2}{s-2} + 1$, there exists a pebbling algorithm for the line of length $n$ that uses at most $s$ pebbles and $O(mn)$ steps.
\end{theorem}
Our \cref{thm-lower-bound} will show that the above algorithm for spooky pebbling the line is asymptotically optimal for all pebble bounds.

\begin{corollary}\label{cor-lin-time-pebbling}
For any constant $\epsilon \in (0,1]$ there exists a spooky pebbling algorithm on \edit{the}{any} line of length $n\edit{}{\geq \ceil{2/\epsilon}^{\ceil{2/\epsilon}}}$ that uses \edit{}{$\ceil{n^\epsilon}$ pebbles and} $O(n/\epsilon)$ steps.
\end{corollary}
\begin{proof}
Let $m$ be the smallest integer such that $n \leq \binom{m + \edit{n^\epsilon}{\ceil{n^\epsilon}} -2}{\edit{n^\epsilon}{\ceil{n^\epsilon}} -2}\edit{}{+1}$. Applying the algorithm in \cref{thm-spooky-pebble-alg} to the line of length $n$ lets us pebble it with $O(mn)$ steps \edit{}{and $\ceil{n^\epsilon}$ pebbles}. We now want to show that when $m' = \edit{2/\epsilon}{\ceil{2/\epsilon}}$ and $n \geq \edit{(2/\epsilon)^{2/\epsilon}}{\ceil{2/\epsilon}^{\ceil{2/\epsilon}}}$, $n \leq \binom{m' + \edit{n^{\epsilon}}{\ceil{n^{\epsilon}}} - 2}{\edit{n^\epsilon}{\ceil{n^\epsilon}} -2}$. Observe that:
\begin{align*}
    \binom{m' + \edit{n^\epsilon}{\ceil{n^\epsilon}} -2}{\edit{n^\epsilon}{\ceil{n^\epsilon}} -2} \geq \left(\frac{m'+n^\epsilon-2}{m'}\right)^{m'} \geq n^2/\edit{(2/\epsilon)^{2/\epsilon}}{\ceil{(2/\epsilon)}^{\ceil{2/\epsilon}}} \geq n
\end{align*}
By our choice of $m$ we can conclude that $m \leq m'$ and therefore the number of steps for the pebbling algorithm is $O(n/\epsilon)$.
\end{proof}
\begin{corollary}\label{cor-low-pebbles-pebbling}
For any $\epsilon \in (0,1]$, there exists a spooky pebbling algorithm on the line of length $n$ that uses $O(1 / \epsilon)$ pebbles and $O(n^{1 + \epsilon} / \epsilon)$ steps.
\end{corollary}
\begin{proof}
Let $m = \edit{(n^\epsilon - 1) / \epsilon}{\ceil{(2n^\epsilon - 1) / \epsilon}}$ and $s = \edit{1 / \epsilon + 2}{\ceil{1 / \epsilon} + 2}$.
\begin{align*}
    \binom{m + s - 2}{s - 2} \edit{}{+1}  \geq \left(\frac{m + s - 2}{s - 2}\right)^{s - 2} = \left(\frac{\edit{(n^\epsilon - 1) / \epsilon}{\ceil{(2n^\epsilon - 1) / \epsilon}} + \edit{1/\epsilon}{\ceil{1 / \epsilon}}}{\edit{1 / \epsilon}{\ceil{1/\epsilon}}}\right)^{\edit{1 / \epsilon}{\ceil{1/\epsilon}}} \edit{=}{\geq \left( \frac{2n^\epsilon / \epsilon}{2/\epsilon}\right) ^{\ceil{1/\epsilon}} \geq}  n
\end{align*}
Therefore, by \cref{thm-spooky-pebble-alg}, there exists a pebbling algorithm on the line using $\edit{O(s)}{s} = O(1 / \epsilon)$ pebbles and \edit{$O(mn) = O(n^{1 + \epsilon} / \epsilon)$}{$O(mn)$ or $O(n^{1+\epsilon}/\epsilon)$} steps.
\end{proof}
\begin{corollary}\label{cor-log-pebbles-pebbling}
There exists a spooky pebbling algorithm on the \edit{line}{lines} of length $n$ that uses $O(\log n)$ pebbles and $O(n \log n)$ steps.
\end{corollary}
\begin{proof}
By \cref{cor-low-pebbles-pebbling} with $\epsilon = 1 / \log n$.
\end{proof}
These results are summarized in \cref{tab-line-pebblings}.

%% file: 4-lower-bounds.tex
\section{Lower bounds on spooky pebbling the line}\label{sec-lower-bounds}

We use a divide and conquer approach to describe the general structure of time-optimal spooky pebbling algorithms for a given pebble bound.
We then analyze the runtime of such an algorithm to obtain an asymptotically tight lower bound on spooky pebbling the line.

\subsection*{Optimal structured spooky pebbling}\label{sec-spooky-structure}

Let $\calP = [\calP_0, \ldots, \calP_\tau]$ be an arbitrary spooky pebbling of the line with nodes $v_1, \ldots v_n$. We will show that $\calP$ uses at least as many steps as a spooky pebbling algorithm matching $\calP'$ in \cref{lem-spooky-pebble-structure}.
\begin{lemma}\label{lem-spooky-pebble-structure}
Let $\calP$ be an arbitrary spooky pebbling of the line using $T(\calP) = \tau$ steps and $S(\calP) = s$ pebbles. Then there exists $k$, 
pebbling algorithm $\calP^*$, and unpebbling algorithm $\calU^*$ such that the pebbling algorithm $\calP'$ described below has $T(\calP') \leq \tau$ and $S(\calP') \leq s$.
\end{lemma}
\begin{lstlisting}
$\calP'(\{v_1, \ldots, v_n\})$:
    place($v_1$)
    for $i \in [2,k]$:
        place($v_{i}$)
        remove($v_{i-1}$)
    run $\calP^*(\{v_{k+1}, \ldots v_n\})$
    run $\calU^*(\{v_1, \ldots, v_k\})$
\end{lstlisting}

In other words, $\calP'$ is the same algorithm as in \cref{lem-spooky-pebble-alg}, but with the choice of $k$ determined by the values of $n$ and $s$ in some arbitrary way.
We prove \Cref{lem-spooky-pebble-structure}
over the course of \cref{sec-spooky-structure}. 

Given a spooky pebbling algorithm and a node $v_i$, we define a notation for the first and last time that this node contains a pebble:
\begin{definition}
Let $\text{First}(v_i,\calP) = \min(\{t | v_i \in \calP_t\})$ and $\text{Last}(v_i, \calP) = \max(\{t | v_i \in \calP_t\})$.
\end{definition}
\begin{lemma}\label{lem-right-to-left-remove} 
$\forall i \in [2,n-1]$, $\text{Last}(v_i, \calP) < \text{Last}(v_{i-1}, \calP)$.
\end{lemma}
\begin{proof}
Let $t = \text{Last}(v_i, \calP)$, which implies that there is never a pebble placed on $v_i$ after $\calP_t$. By the definition of a spooky pebbling we require that $\calG_\tau = \emptyset$, so $v_i \not \in \calG_{t+1}$ as we never place a pebble on $v_i$ after step $t$. Since this is a valid pebbling, we can conclude that $v_{i-1} \in \calP_t$ and thus $\text{Last}(v_i, \calP) < \text{Last}(v_{i-1}, \calP)$.
\end{proof}
\begin{definition}
Let $\text{First}(v_n,\calP) = t$. We call $\calP$ \emph{sweep-first} if a pebble is placed at $v_n$ only once and each $v_i$ has at most one pebble placed on it before step $t$.
\end{definition}

\begin{lemma}\label{lem-sweep-emulate-optimal}
    Let $G = (V,E)$ be a line and $S \subseteq V$ be such that $|S| = s$. Then there exists a valid sequence of pebbling instructions that places pebbles on exactly $S$ that is time-optimal and uses at most $s+1$ pebbles.
\end{lemma}
\begin{proof}
Consider the following pebbling sequence:
\begin{lstlisting}
$\calI(\{v_1, \ldots, v_n\}, S)$:
    $k = \text{argmax}_{i} v_i \in S$
    place($v_1$)
    for $i \in [2,k]$:
        place($v_{i}$)
        if($v_{i-1} \not \in S$):
            remove($v_{i-1}$)
\end{lstlisting}
The above sequence uses at most $s+1$ pebbles and takes exactly $2k-s$ steps. Since placing a pebble on $v_k$ requires placing pebbles on each node before it and we want to only end with pebbles on $S$, any pebbling sequence that completes this task must place at least $k$ pebbles and remove $k-s$ pebbles. Therefore the above pebbling sequence uses the time-optimal number of steps.
\end{proof}

Note that $\calI$ only uses $s$ pebbles when $v_{k-1} \in S$. Using the above lemma, we can turn any pebbling into one that is sweep first.

\begin{lemma}\label{lem-sweep-first}
Let $\calP$ be any spooky pebbling algorithm where $T(\calP) = \tau$ and $S(\calP) = s$. Then there exists a sweep-first pebbling algorithm $\calP'$ where $T(\calP') \leq \tau$ and $S(\calP') \leq s$.
\end{lemma}
\begin{proof}
Let $t$ be the last step where $\calP$ places a pebble on $v_n$.
We know that $t < \text{Last}(v_{n-1}, \calP)$. Thus by \Cref{lem-right-to-left-remove} we know that $t < \text{Last}(v_i, \calP)$ for all $i < n$.
$\calP'$ will place pebbles on exactly $\calP_t$ in $t' = 2n - \norm{\calP_t}$ steps by running $\calI(\{v_1, \ldots v_n\}, \calP_t)$. 
We know that $t' \leq t$ since each $v_i$ has a pebble placed and each node not in $\calP_t$ must have had a pebble  removed before step $t$ in $\calP$. We define $\calP' = \left[\calP_0, \calP'_1, \ldots, \calP'_{t'} = \calP_t, \calP_{t+1}, \ldots, \calP_{\tau} \right]$. Since $t' \leq t$ we know that $T(\calP') \leq \tau$. We can additionally conclude that $S(\calP') \leq s$ since $\calP_t$ must contain a pebble at $v_{n-1}$ and so $\calI$ only uses $s$ pebbles. Note that while $\calP$ and $\calP'$ may have different ghosting sequences, $\calP'$ is still a valid spooky pebbling. This is because $\calP$ must place a pebble (and subsequently remove it without creating a ghost) at every location where $\calP'$ had a ghost during step $t'$ due to \cref{lem-right-to-left-remove}.
\end{proof}
Now we are ready to prove \cref{lem-spooky-pebble-structure}.
\begin{proof}[Proof of \cref{lem-spooky-pebble-structure}]
By \cref{lem-sweep-first} we can assume that $\calP$ is sweep-first. Without loss of generality, let $\calP$ be the algorithm with $T(\calP) = \tau$ and $S(\calP) = s$ that maximizes over all algorithms the least value of $k$ where $v_k \in \calP_{\text{First}(v_n, \calP)}$.
Let $t = \text{First}(v_n, \calP)$.
By our choice of $\calP$ and $k$, for all $i < k$, $\calP$ must place and remove a pebble at $v_i$;
thus we can rearrange these steps from $\calP$ to match lines 2 through 5 of $\calP'$.
Since $\calP$ is a sweep-first pebbling we know that these are the only instructions acting on $v_1, \ldots, v_k$ before step $t$.
We now partition the remaining instructions of $\calP$ into $\calP^*$, the set of all instructions acting on nodes $v_{k+1}, \ldots, v_n$, and $\calU^*$, the set of all instructions acting on $v_{1}, \ldots, v_k$ --- while maintaining their relative order within $\calP$. 

We will now show that $\calP^*$ is a valid pebbling of $\{v_{k+1}, \ldots, v_n\}$ that uses at most $s-1$ pebbles.
Since (i) $\calP$ is a valid pebbling, (ii) $\calP'$ as constructed above maintains a pebble at $v_k$ during the execution of $\calP^*$, and (iii) instructions acting on vertices $v_1, \ldots, v_{k-1}$ cannot change the validity of an instruction on $v_{k+1}, \ldots, v_n$, it follows that $\calP^*$ is a valid pebbling.
Now assume for the sake of contradiction that $\calP^*$ placed $s$ pebbles during some step.
Then since $\calP$ uses only $s$ pebbles, there must be some first step $t'$ in $\calP$ where $\{v_1, \ldots, v_k\} \cap \calP_{t'} = \emptyset$ before the $s$'th pebble is placed on vertices $\{v_{k+1}, \ldots, v_n\}$.
Let $\ell$ be the least value where $v_\ell \in \calP_{t'}$. 
We will construct another sweep-first pebbling algorithm $\calP^\perp$ that has $\ell > k$ as the least value where $v_\ell \in \calP^\perp_{\text{First}(v_n, \calP^\perp)}$ while maintaining $T(\calP^\perp) \leq \tau$ and $S(\calP^{\perp}) \leq s$, contradicting our choice of $\calP$.
Intuitively $\calP^\perp$ will behave like $\calP$ except that it has a pebble at $v_\ell$ instead of $v_k$ during step $t$.
$\calP^\perp$ has $\calP^\perp_{\text{First}(v_n, \calP^\perp)} = (\calP_t \cup \{v_\ell\}) \setminus \{v_k\}$ and reaches this state by following the first $t$ instructions of $\calP$, except that $\calP^\perp$ ghosts $v_k$ after a pebble is placed on $v_{k+1}$ and ignores any instruction to remove $v_\ell$ before placing a pebble on $v_n$. 

After this, $\calP^\perp$ continues following the instructions of $\calP$ on all vertices excluding $v_k, \ldots v_\ell$ until step $t'$ of $\calP$.
At this point we know that $\calP_{t'}$ has empty intersection with $v_k, \ldots, v_{\ell-1}$ and therefore $\calP$ and $\calP^\perp$ have pebbles in the same positions and ghosts on the same positions excluding $v_k, \ldots, v_{\ell-1}$. 
However $\calP_{t'}$ has a pebble at $v_\ell$ and therefore must place pebbles on all of $v_k, \ldots, v_{\ell-1}$.
So if $\calP^\perp$ copies all steps of $\calP$ after step $t'$, it will also remove any ghosts it had in $v_{k}, \ldots v_{\ell -1}$.
Thus $\calP^\perp$ is a valid spooky pebbling that uses at most $s$ pebbles and $\tau$ steps and violates our choice of $\calP$ since $\ell > k$.

We know that $\calU^*$ must be a valid unpebbling by the same reasoning as $\calP^*$.
All that remains is to show that $\calU^*$ uses at most $s-1$ pebbles.
As $\calP$ is a sweep-first algorithm that uses at most $s$ pebbles, we know that for all steps $t' > t$, $\norm{\calP_{t'} \cap \{v_1, \ldots, v_k\}} \leq s-1$.
Thus executing the same instructions in $\calU^*$ cannot use more than $s-1$ pebbles and $\calP'$ is a valid spooky pebbling that uses no more pebbles or steps than $\calP$.
\end{proof}

By \cref{lem-spooky-pebble-structure} we can assume that for any fixed pebble count, there is a minimum step algorithm that matches the structure of $\calP'$. By recursively applying this argument, we get a recurrence relation for the minimum number of steps needed to pebble the line of length $n$ using at most $s$ pebbles. In general this recurrence has a tree like structure based on the choice of $k$ at each level.
However, the value of $k$ that minimizes the expression is the optimal choice.

\begin{corollary}\label{cor-pebbling-formula}
Let $T_P(n,s)$ ($T_{UP}(n,s)$) be the minimum number of steps needed to pebble (unpebble) a line of length $n$ using at most $s$ pebbles. Then:
\begin{align*}
    T_P(n,s) = \begin{cases}
                    1 & s \geq 1 \text{ and } n=1 \\
                    \infty & s < 3 \text{ and } s < n\\
                    \min_{k \in [1,n)} 2k-1 + T_{UP}(k, s-1) + T_P(n-k,s-1) & \text{o.w.}
                \end{cases}\\
\end{align*}
\end{corollary}
The $\infty$ comes from observing that there is no way to spooky pebble a line longer than $s$ with $s$ pebbles when $s<3$.
Unfortunately the above recurrence features a call to $T_{UP}$ and the non-obvious choice of $k$ makes it hard to analyze.
Now we establish how the time complexities of pebbling, unpebbling, and unghosting relate so that we can convert \cref{cor-pebbling-formula} into a form that is easier to work with.

\subsection*{Relationships between Pebbling, Unpebbling, and Unghosting}

In \cref{sec-line-pebble}, our constructions were more naturally expressed using unghosting rather than pebbling. In order to prove the tightness of our upper bounds, it is convenient to also express our lower bounds in terms of unghosting. For such a bound to be useful, we require a tight relationship between pebbling and unghosting. We have already proven a way to convert an unghosting to a pebbling.

\unghostingtopebbling*

Now we will show how to convert a pebbling into an unghosting.

\begin{lemma}\label{lem-pebbling-to-unghosting}
Let $\calP$ be a pebbling algorithm for the line of length $n+1$ that uses $s+1$ pebbles and takes $\tau+1$ steps. Then there exists an unghosting algorithm on the line of length $n$ that uses $s$ pebbles and takes at most $\tau$ steps.
\end{lemma}
\begin{proof}
Let $t$ be the the last step of $\calP$ operating on $v_{n+1}$. This must be a pebbling step, so both $v_n$ and $v_{n+1}$ are in $\calP_t$. Divide $\calP$ into two components around this point, $\calP_\text{pre}$ for the first $t$ steps and $\calP_\text{post}$ for the remaining steps. Since $\calP_\text{pre}$ must have spent at least one step on pebbling each of $v_n$ and $v_{n+1}$, there must be some algorithm $\calP_\text{pre}^*$ running in at most $t-2$ steps that ends in $\calP_t \setminus \{v_n, v_{n+1}\}$. Moreover, by Lemma~\ref{lem-sweep-emulate-optimal}, $\calP_\text{pre}^*$ may be constructed to require at most $|\calP_t \setminus \{v_n, v_{n+1}\}| + 1 \leq s$ pebbles.

Let $\calP_\text{post}^*$ be a sequence of pebbling instructions derived from $\calP_\text{post}$ by removing any instructions operating on $v_n$ before $\text{Last}(v_n, \calP_\text{post})$, then inserting an instruction placing a pebble on $v_n$ immediately before that last removal of $v_n$. This is valid, as the new placement occurs immediately before a removal (so the predecessor of $v_n$ is pebbled), and no operations occur on $v_{n+1}$ that would be affected by changing whether a pebble is present on $v_n$.

$\calP_\text{pre}^*$ followed by $\calP_\text{post}^*$ is then a valid unghosting of the line graph up to $v_n$, excluding $v_{n+1}$. Since $\calP_\text{pre}^*$ is at least 2 steps shorter than $\calP_\text{pre}$ and $\calP_\text{post}^*$ is at most 1 step longer than $\calP_\text{post}$, the composition takes at most $\tau$ steps. Moreover, the composition requires at most $s$ pebbles, $\calP_\text{pre}^*$ by Lemma~\ref{lem-sweep-emulate-optimal} and $\calP_\text{post}^*$ because it operates as in $\calP_\text{post}$ but without an extra pebble sitting on $v_{n+1}$.
\end{proof}

Combining the results of \cref{lem-pebbling-to-unghosting} and \cref{lem-unghosting-to-pebbling} we show a tight relationship between unghosting and pebbling. We also characterize the relationship between unghosting and unpebbling.

\begin{lemma}
Let $\calU_g$ be an unghosting algorithm for the line of length $n$ that uses $s$ pebbles and takes $\tau$ steps. Then there exists an unpebbling algorithm for the line of length $n$ that uses $s$ pebbles and takes $\tau+1$ steps.
\end{lemma}
\begin{proof}
Define $\calU_p$ by first ghosting $v_n$, then running $\calU_g$.
\end{proof}

\begin{lemma}
Let $\calU_p$ be an unpebbling algorithm for the line of length $n$ that uses $s$ pebbles and takes $\tau$ steps. Then there exists an unghosting algorithm for the line of length $n$ that uses $s$ pebbles and takes $\tau+1$ steps.
\end{lemma}
\begin{proof}
Define $\calU_g$ by running $\calU_p$, but before $\text{Last}(v_n, \calU_p)$, skip all preexisting operations on $v_n$, and immediately before $\text{Last}(v_n, \calU_p)$, add a new step to place a pebble on $v_n$. $\calU_g$ only differs from $\calU_p$ by sometimes not having a pebble on $v_n$ when $\calU_p$ does, so it uses at most as many pebbles, and it inserts only a single new step.
\end{proof}
\begin{theorem}\label{thm-pebble-unpebble-unghost}
$T_P(n+1, s+1) = T_{UG}(n, s) + 1$. $T_{UG}(n, s) = T_{UP}(n, s) \pm 1$.
\end{theorem}

\subsection*{Lower bounds on the number of steps for pebbling}\label{sec-pebble-lb}
Now that we have established the relationship between pebbling, unpebbling, and unghosting, we can lower bound the number of steps needed to perform these tasks.
While pebbling is the most natural quantity to evaluate, it turns out that unghosting has a much cleaner analysis than pebbling.

We can now restate \cref{cor-pebbling-formula} in the language of unghosting.
\begin{corollary}\label{cor-recurrance-unghosting}
Let $T_{UG}(n,s)$ ($T_{UP}(n,s)$) be the minimum number of steps needed to unghost (unpebble) a line of length $n$ using at most $s$ pebbles. Then:
\begin{align*}
    T_{UG}(n,s) = \begin{cases}
                    2 & s \geq 1 \text{ and } n = 1 \\
                    \infty & s < 2 \text{ and } s < n\\
                    \min_{k \in [1,n]} 2k -1 + T_{UP}(k, s) + T_{UG}(n-k,s-1) & \text{o.w.}
                \end{cases}\\
\end{align*}
\end{corollary}

Note that we cannot apply \cref{thm-pebble-unpebble-unghost} to the base case in \cref{cor-pebbling-formula}. Instead the base case in the above recurrence can be verified by inspection. We can then use \cref{thm-pebble-unpebble-unghost} to convert our lower bound on unghosting to pebbling.
\edit{In the rest of this section, we prove that unghosting with at most $s$ pebbles requires $\Omega(mn)$ steps.}{}
By \cref{thm-pebble-unpebble-unghost} we know that $T_{UP}(k,s) \geq T_{UG}(k,s) -1$. We therefore define the following recurrence relation that is a lower bound on $T_{UG}(n,s)$:

\begin{equation}\label{eqn-time-recur}
    T(n,s)= \begin{cases}
                    2 & s \geq 1 \text{ and } n=1 \\
                    \infty & s < 2 \text{ and } s < n\\
                    \min_{k \in [1,n]} 2k-2 + T(k, s) + T(n-k,s-1) & \text{o.w.}
                \end{cases}\\
\end{equation}
\edit{}{While $T(n,s)$ is not exactly the minimum number of steps needed to unghost the line of length $n$ with $s$ pebbles, it is a sufficiently tight lower bound.}
A \edit{similar function}{function similar to $T(n,s)$} appears in \cite{WH00, New08} as 
\edit{a}{the} solution to the backtracking problem in dynamic programming. They provide efficient algorithms that can be used to compute the optimal value of $k$ in each recursive call. Another similar recurrence appears in the problem of sequence reversing, where there is a proven lower bound similar to the one we present here \cite{Pott95, GPRS97}.
\edit{[Move this point elsewhere:]
While $T(n,s)$ is not exactly the minimum number of steps needed to unghost the line of length $n$ with $s$ pebbles, it is a sufficiently tight lower bound.}{In the rest of this section we use this recurrence relation to prove that unghosting with at most $s$ pebbles requires $\Omega(mn)$ steps which is sufficient to obtain the following theorem.}

%% file: 4.3-tree-lb-proof.tex
\begin{theorem}\label{thm-lower-bound}
For any line of length $n$, any pebble bound $s$, and any positive integer $m$ where $n \geq \binom{m + s-2}{s-2} + 1$, a spooky pebbling of that line with at most $s$ pebbles uses at least $\Omega(mn)$ steps.
\end{theorem}

One difficulty in analyzing \edit{the recurrence above}{our recurrence relation} is that the choice of $k$ that minimizes this expression is not obvious, making it difficult to directly compute this function.
Instead we will describe a family of related recurrence relations for a fixed choice of $n$ and $s$ using trees with $n$ leaf nodes and right-depth $s-1$ to determine the choices of $k$. The lowest valued recurrence in this family gives us the value of $T(n,s)$.

\begin{definition}
Let $\calT$ be a binary tree, $\calT.l$ be the left subtree, and $\calT.r$ be the right subtree.
Let $|\calT|_L$ be the number of leaves in a given binary tree.
If $|\calT|_L = n$ then we can define the recurrence relation:
\begin{equation}
    T_\calT(n,s) = \begin{cases}
        2 & s\geq 1 \text{ and } n = 1 \\
        \infty & s < 2 \text{ and } s < n \\
        2|\calT.l|_L-2 + T_{\calT.l}(|\calT.l|_L,s) + T_{\calT.r}(|\calT.r|_L, s-1) & \text{o.w.}
    \end{cases}
\end{equation}
\end{definition}

We can think of this as being \cref{eqn-time-recur} except that $k$ is selected as the number of leaves in the left sub-tree rather than the value that minimizes the expression.
This lets us fix $n$ and $s$ in order to see how different choices of $k$ lead to different values for the recurrence. Then the minimum of $T_\calT (n,s)$ over all trees must equal to $T(n,s)$. Instead of directly computing the recurrence, we can compute $T_\calT(n,s)$ for a specific tree $\calT$ by assigning a cost function to binary trees that is equal to $T_\calT(n,s)$.

\begin{definition}
Let $N$ be any node in a tree. We let $R(N)$ denote the right-depth of $N$, which is equal to the number of times you must take right children to reach $N$ from the root. We likewise define $L(N)$ as the left-depth of $N$.
\end{definition}
\begin{definition}
    For any depth bound $s$, the \emph{cost} of a binary tree $C_s(\calT)$ is equal to the sum of the costs of its nodes. Each non-leaf node of a tree has cost $-2$. Any leaf node $N$ where $R(N) < s$ has a cost of $2L(N) + 2$ while a leaf node where $R(N) \geq s$ has a cost of $\infty$.
\end{definition}


\begin{lemma}\label{lem-tree-pebble-cost-equiv}
The cost of the tree $C_s(\calT)$ is equal to $T_\calT(|\calT|_L,s)$.
\end{lemma}
\begin{proof}
We prove this by structural induction on the tree. In the base case we have a tree that is a single leaf. Then $C_s(\calT) = T_\calT(n,s) = 2$ as desired.
Now consider an arbitrary tree $\calT$. By the inductive hypothesis we can assume that $C_s(\calT.l) = T_{\calT.l}(|\calT.l|_L, s)$ and $C_{s-1}(\calT.r) = T_{\calT.r}(|\calT.r|_L, s-1)$. Given that the cost of a tree is the sum of the costs of its nodes, $C_s(\calT) = 2|\calT.l|_L - 2 \edit{}{+} C_s(\calT.l) + C_{s-1}(\calT.r)$, as all nodes in $\calT.l$ have their left depth increased by one, all nodes in $\calT.r$ has their right depth increased by one, and the root of $\calT$ has a cost of $-2$. By our inductive hypothesis, this implies that
\begin{equation*}
    C_s(\calT) = 2 |\calT.l|_L -2 + T_{\calT.l}(|\calT.l|_L, s) + T_{\calT.r}(|\calT.r|_L, s-1) = T_{\calT}(|\calT|_L, s).
\end{equation*}
If we have a tree with a leaf node where $R(N) \geq s$ then the tree has a cost of infinity and $T_\calT(n,s)$ must reach the base case where $s \leq 1$ and $s < n$ when evaluating that node.
\end{proof}

\begin{corollary}\label{cor-min-cost-tree-name}
    Let $T_{(n,s)}$ be the tree with $n$ leaf nodes that minimizes $C_s(T_{(n,s)})$. Then $T(n,s) = C_s(T_{(n,s)})$.
\end{corollary}

It is easier to reason about $C_s(T_{(n,s)})$ than $T_{\edit{T_(n,s)}{}}(n,s)$ as we can start with an arbitrary tree with $n$ leaf nodes and show how it would be possible to mutate that tree and reduce its cost without computing the full value of the recurrence relation.

Now that we have established the equivalence between the cost of a tree $T_{(n,s)}$ and $T(n,s)$, we want to prove some things about the structure of this tree. When $s<2$ and $s<n$ we know that $T(n,s) = \infty$, so whenever possible a tree will never have a sub-tree with such parameters as nodes. This means that any tree $T_{(n,2)}$ must have $T_{(1,1)}$ as its right child. By construction, this means that the right-depth of any node in the tree $T_{(n,s)}$ is at most $s-1$. Note that the cost of all the intermediate nodes is only a function of the total number of nodes in the tree.

What follows is a collection of technical lemmas that give the number of nodes in $T_{(n,s)}$ with specific left and right depths. Together they let us characterize the number of leaves in $T_{(n,s)}$ with each cost. This lets us construct a lower bound on the cost of $T_{(n,s)}$, which in turn is a lower bound on the total number of steps needed to unghost a line of length $n$ using at most $s$ pebbles.

\begin{lemma}\label{lem-max-cost-leaf}
Let $T_{(n,s)}$ be the tree in \cref{cor-min-cost-tree-name}. Let $X$ be a leaf of $T_{(n,s)}$ with the largest value for $L(X)$, $m=L(X)$, and $X'$ be any other leaf where $L(X') < m-1$. 
Then $R(X') = s-1$.
\end{lemma}
\begin{proof}
Assume that $R(X') < s-1$.
Note that $X$ must be a left child, as it is the node with largest left depth.
Then we could replace the parent of $X$ with its other child and replace $X'$ with an intermediate node whose left child is $X$ and right child is $X'$.
Doing so must reduce the left depth of $X$ by at least one and does not change the left depth of any other node in the tree, so the new tree has a lower cost. 
But this is a contradiction since $T_{(n,s)}$ is supposed to be the tree with the lowest cost.
\end{proof}

The above lemma tells us a surprising amount about the structure of $T_{(n,s)}$.
Since all nodes $X$ where $L(x) < m-1$ and $R(X) < s-1$ cannot be leaves, these nodes all have left and right children.
We can therefore derive exactly how many leaves $T_{(n,s)}$ must have with each left depth less than $m$.
\begin{lemma}\label{lem-leaf-counts}
    Let $T_{(n,s)}$ be the tree in \cref{cor-min-cost-tree-name} and $m$ be the largest value of $L(N)$ for $N \in T_{(n,s)}$.
    If $m>1$, then we know that the number of leaf nodes $T_{(n,s)}$ possesses with left depth $L<m$ is $\binom{L+s-2}{s-2}$.
\end{lemma}
\begin{proof}
    By \cref{lem-max-cost-leaf} we know that no node with a left depth less than $m-1$ and a right depth less than $s-1$ can be a leaf.
    Thus all such nodes must have left and right children.
    We will identify each node with a string of $l$'s and $r$'s indicating which directions are taken from the root of the tree to arrive at that node.
    Since all nodes with right depth $s$ or larger have cost infinity,
    we know that all nodes with right depth $s-1$ must be leaf nodes.
    Thus each leaf node with left depth less than $m-1$ must be a right child (otherwise, their sibling would have right depth $s$) and be identified with a string ending in an $r$.
    There are $\binom{L+s-2}{s-2}$ strings containing $L$ copies of $l$ and $s-1$ copies of $r$ that end in an $r$. This exactly corresponds to the number of leaf nodes with left depth $L < m-1$; each of these leaf nodes must exist in the tree because none of their ancestors can be leaves by \cref{lem-max-cost-leaf}, and there can be no other leaves with this left depth unless there is a leaf with cost $\infty$.
    
    Since, by \cref{lem-max-cost-leaf}, all nodes with left depth $L = m-2$ and right depth $R < s-1$ must exist in the tree and cannot be leaves, each of these nodes has a left child.
    There is a bijection between the number of leaves with left depth $L=m-1$ and these children, as their rightmost descendants are exactly the leaf nodes with left depth $m-1$.
    Since there are $\binom{m+R-2}{R}$ nodes with left depth $m-2$ and right depth $R < s-1$, there are
    \begin{equation*}
        \sum_{R = 0}^{s-2} \binom{m + R - 2}{R} = \sum_{R=0}^{s-2} \binom{m+R-2}{m-2} = \binom{m+s-3}{m-1} = \binom{L + s - 2}{s-2}
    \end{equation*}
    leaf nodes with left depth $m-1$.
\end{proof}

\begin{lemma}\label{lem-monotonic-left-depth}
Let $T_{(n,s)}$ be the tree in \cref{cor-min-cost-tree-name} and $m$ be the largest left depth of any node in $T_{(n,s)}$ and $m'$ be the largest left-depth of any node in $T_{(n',s)}$. If $m' > m$ then $n'> n$.
\end{lemma}
\begin{proof}
Assume that $n' \leq n$. By \cref{lem-leaf-counts} for all $L < m$, $T_{(n,s)}$ and $T_{(n',s)}$ must have the same number of leaf nodes with left depth $L$. This means that $T_{(n,s)}$ must have at least $n-n'$ more leaves with left depth $m$ than $T_{(n',s)}$ has with left depth at least $m$. We construct a new tree $T^*_{(n',s)}$ by taking $T_{(n,s)}$ and removing $n-n'$ leaves with left depth $m$ by replacing their parent with its right child. The tree $T^*_{(n',s)}$ represents a valid binary tree and its cost is less than that of $T_{(n',s)}$ since the two trees have the same total number of nodes, the same number of leaves with any left depth less than $m$, and $T^*_{(n',s)}$ has no leaves with left depth larger than $m$. This implies that $T_{(n',s)}$ was not constructed according to \cref{cor-min-cost-tree-name}, giving us a contradiction.
\end{proof}

\begin{lemma}\label{lem-leaves-depth-m}
Let $T_{(n,s)}$ be the tree in \cref{cor-min-cost-tree-name} and $n$ be the smallest value such that $T_{(n,s)}$ contains a node with left depth $m+1$. Then $T_{(n-1,s)}$ contains exactly $\binom{m+s-2}{s-2}$ leaves with left depth $m$ and $n =1 + \binom{m + s-1}{s-1}$.
\end{lemma}
\begin{proof}
Assume for the sake of contradiction that $T_{(n,s)}$ contained at least two nodes with left depth $m+1$. By our choice of $n$ we know that $T_{(n-1,s)}$ contains no nodes with left depth $m+1$. By \cref{lem-leaf-counts} we know that $T_{(n,s)}$ and $T_{(n-1,s)}$ have the same number of leaves with each left depth less than $m$. This means the difference in cost between $T_{(n,s)}$ and $T_{(n-1,s)}$ must be at least $m$.\footnote{$T_{(n,s)}$ must have two leaves with left depth $m+1$ while $T_{(n-1,s)}$ has one fewer node that must have left depth $m$. Since $T_{(n,s)}$ has one more intermediate node, its cost is at least $m$ more than $T_{(n-1,s)}$.} But there is another tree $T^*_{(n,s)}$ that costs only $m-1$ more than $T_{(n-1,s)}$ constructed by replacing the leftmost leaf of $T_{(n-1,s)}$ with an intermediate node whose children are both leaves. Since $T^*_{(n,s)}$ has cost less than $T_{(n,s)}$, we get a contradiction.

Now $T_{(n,s)}$ must contain exactly one node with left depth $m+1$. By \cref{lem-leaf-counts} this means that it contains exactly $\binom{L + s -2}{s-2}$ leaves with left depth $L$, so the total number of leaves in $T_{(n,s)}$ is:
\begin{align*}
    n &= 1 + \sum_{L=0}^{m} \binom{L + s -2}{ s -2} \\
    &=1 + \binom{m + s-1}{s-1}
\end{align*}
This means that $T_{(n-1,s)}$ must have $\binom{m+s-1}{s-1}$ leaves. All of these leaves have left depth at most $m$ by \cref{lem-monotonic-left-depth}. So \cref{lem-leaf-counts} gives us that $T_{(n-1,s)}$ has:
\begin{align*}
    &\binom{m+s-1}{s-1} - \left( \sum_{L=0}^{m-1} \binom{L+s-2}{s-2} \right)\\
    &= \binom{m + s -1}{s-1} - \binom{m + s -2}{s-1}\\
    &= \binom{m+s-2}{s-2}
\end{align*}
leaves with left depth $m$.
\end{proof}

\begin{lemma}\label{lem-tree-cost-lb}
Let $T_{(n,s)}$ be the tree in \cref{cor-min-cost-tree-name} and $\binom{m + s-1}{s-1} \leq n$. Then the cost of $T_{(n,s)}$ is $\Omega(mn)$.
\end{lemma}
\begin{proof}
By \cref{lem-leaf-counts}, \cref{lem-monotonic-left-depth}, and \cref{lem-leaves-depth-m}, we know that for all $L \in \{0,1,2, \ldots, m\}$ the number of leaf nodes with left depth $L$ is $\binom{L + s-2}{s-2}$. Thus the cost of these leaf nodes is at least:
\begin{align*}
    \sum_{L = 0}^{m} (2L+2) \binom{L + s -2}{s-2}.
\end{align*}
Expanding the above summation yields:
\begin{equation*}
    \frac{2 (m+1) (m(s-1) + s)}{s(s-1)} \binom{m+s-1}{s-2}.
\end{equation*}
Using that $\binom{n}{k-1} = \frac{k}{n-k+1} \binom{n}{k}$ and $s\geq 2$, this is at least:
\begin{align*}
    \frac{2(m(s-1)+s)}{s} \binom{m+s-1}{s-1} &= 2\left(\frac{m(s-1)}{s} + 1\right) \binom{m+s-1}{s-1}\\
    &\geq (m+2)\binom{m+s-1}{s-1}.
\end{align*}
By \cref{lem-leaf-counts} and \cref{lem-leaves-depth-m}, if $n=\binom{m + s-1}{s-1} + k$ then the remaining $k$ leaves must all have left depth larger than $m$.
Thus the total cost of the leaves is at least:
\begin{align*}
    (m+2) \binom{m+s-1}{s-1} + 2(m+1)k \geq  (m+2)n.
\end{align*}
There are $n-1$ intermediate nodes with cost $-2$ so the total cost of the tree must be at least:
\begin{align*}
    (m+2)n -2(n-1) = mn-2.
\end{align*}
Which is $\Omega(mn)$ as desired.
\end{proof}
By applying \cref{lem-tree-cost-lb} to lower bound the cost of trees and \cref{lem-tree-pebble-cost-equiv} to apply this lower bound to \cref{eqn-time-recur}, we obtain the following corollary:
\begin{corollary}\label{cor-unghost-lb}
The number of steps needed to unghost a line of length $n \geq \binom{m+s-1}{s-1}$ with $s$ pebbles is $\Omega(mn)$.
\end{corollary}
By applying \cref{thm-pebble-unpebble-unghost} to \cref{cor-unghost-lb}, we get \cref{thm-lower-bound}.

%% file: 6-DAG-spooky-pebble.tex
\section{Spooky pebbling \edit{more general DAGs}{beyond the line}}
In \cref{sec-line-pebble} and \cref{sec-lower-bounds} we discussed how the spooky pebble game can be applied to simulate arbitrary irreversible sequential computation on a quantum computer.
While this gives us very general results, additional structure on the classical computation can lead to more efficient simulation.
Here we consider computation that can be expressed using a dependency graph---such as boolean circuits, straight line \edit{}{(or oblivious)} programs, and data-independent memory hard functions \edit{}{(e.g. \cite{WT77, Pip78, LT79, Tom80, AS15, RD16, ABP17})}.
When considering a pebbling of a DAG, each pebble represents only a single word rather than an entire copy of the computation's state.
It turns out that there is a close relationship between the number of pebbles required in the irreversible and spooky pebble games.
\subsection*{Spooky pebbling from irreversible pebbling}
\edit{}{We start by showing a general strategy that can be used to convert any irreversible pebbling into a spooky pebbling that uses at most one additional pebble. This result is a special case of Theorem 1 in \cite{QL23, QL24} which presents a generalization of this argument for DAGs with multiple target nodes (nodes with out-degree zero).
For completeness we show the proof below.}
\begin{lemma}\label{lem-irreversible-spooky-pebble-bound}
If a DAG $G$ with $n$ nodes \edit{}{and one target }can be pebbled in the irreversible pebble game using $T$ steps and $s$ pebbles, then it can be spooky pebbled using $O(nT)$ steps and at most $s + 1$ pebbles.
\end{lemma}
\begin{proof}
Let $\calP$ be an irreversible pebbling of $G$ that uses $s$ pebbles. We will construct the following spooky pebbling $\calP_s$ that uses at most $s+1$ pebbles. $\calP_s$ first simulates $\calP$, which since we are operating in the spooky pebble game, may leave some ghosts behind. Additionally the first time that $\calP_s$ places a pebble at any node $v_i$, we push $v_i$ onto an initially empty stack. Once $\calP_s$ has placed a pebble at the target node $v_t$, it removes all other pebbles (leaving a large number of ghosts) and repeats the following procedure until the stack is empty: We pop the top element off of our stack and call it $v_g$. If $v_g$ does not contain a ghost, we continue to the next element in the stack. Once we find a $v_g$ that contains a ghost, we have $\calP_s$ simulate $\calP$ until $v_g$ is pebbled again, leaving ghosts on some nodes. $\calP_s$ then removes the pebble at $v_g$ (without creating a ghost) and then removes all pebbles other than $v_t$, possibly addling ghosts. By our choice of $v_g$, future iterations of this procedure cannot result in a ghost on a prior $v_g$, so this procedure terminates in less than $n$ iterations with pebbles only on $v_t$ and no ghosts. Since $\calP_s$ is replaying steps from $\calP$ with at most one additional pebble on $v_t$, it will use at most $s+1$ pebbles.
\end{proof}
Note that since $T$ is without loss of generality $\Omega(n)$, we can say that the above procedure also runs in time $O(T^2)$.
Since any spooky pebbling is also an irreversible pebbling, the above lemma is sufficient to give us a nearly tight two sided bound on the minimum number of pebbles needed in the spooky pebble game.
\begin{corollary}\label{cor-spooky-uses-one-more-pebble}
Let $G$ be a DAG that requires $s$ pebbles in the irreversible pebble game. The fewest pebbles needed for the spooky pebble game on $G$ is either $s$ or $s+1$.
\end{corollary}
Demaine and Liu proved that determining the fewest pebbles needed in the irreversible pebble game is PSPACE-hard to approximate.
\begin{proposition}[Theorem 1\protect\footnotemark 
 \;in~\cite{DL17}]\footnotetext{\edit{}{Although Theorem 1 as stated in \cite{DL17} does not refer to the number of targets, the full construction presented in section 2.3.2 of \cite{DQ17_full} is a DAG with a single target node.}}\label{thm-irreversible-pebble-hard}
The minimum number of pebbles needed in the irreversible pebble game on DAGs with $n$ nodes\edit{}{, a single target, }and maximum in-degree 2 is PSPACE-hard to approximate to within an additive $n^{1/3 - \epsilon}$ for any $\epsilon > 0$.
\end{proposition}

\noindent
\edit{In \mbox{\cref{cor-spooky-uses-one-more-pebble}} we showed that the number of pebbles needed in the spooky pebble game is closely tied to the same value for the irreversible pebble game.}{}
By combining \cref{thm-irreversible-pebble-hard} with \cref{cor-spooky-uses-one-more-pebble}, we see that finding \edit{}{even approximate } minimum pebble spooky pebbling algorithms \edit{is}{for arbitrary DAGs solves a }PSPACE-hard \edit{}{problem}. Thus unless $P=\text{PSPACE}$, there can be no polynomial time algorithm that generates minimum pebble spooky pebbling strategies on arbitrary DAGs.
\edit{This result was independently proven in \cite{QL24}; however, their proof only holds for graphs with large in-degree and only shows hardness for computing the minimum number of pebbles rather than approximating this quantity.}{Quist and Laarman showed the PSPACE-completeness of this problem in \cite{QL24} through a different reduction;
however, going through \cref{thm-irreversible-pebble-hard} extends this result to a restricted subset of DAGs (those with maximum in-degree two with a single sink node) and shows that even approximating the minimum number of pebbles to an additive factor of $n^{1/3 - \epsilon}$ is PSPACE-hard.}
\begin{theorem}
The minimum number of pebbles needed in the spooky pebble game on DAGs with $n$ nodes\edit{}{, a single sink,} and maximum in-degree 2 is PSPACE-hard to approximate.
\end{theorem}

\subsection*{Spooky pebbling the \edit{}{binary} tree \edit{}{graph}}
\edit{}{While finding the minimum number of pebbles needed in the spooky pebble game is PSPACE-hard to approximate in general, it is possible to find pebble and step efficient strategies for specific graph topologies.}
The binary tree \edit{}{with directed edges from the leaves to the root} is a natural topology for algorithms like divide and conquer, where a problem is solved by combining the results of two smaller sub-problems.
Concrete algorithms with this structure include floating point pairwise summation and computing Merkle trees \cite{CMDJWS64, Mer88}.

\edit{Cook showed that the complete binary tree of height $h$ with $n = 2^h - 1$ nodes requires exactly $h+1$ pebbles and $O(n)$ steps in the irreversible pebble game \cite{Coo73}. 
Kr{\'a}l'ovi{\v{c}} showed that this tree requires $h+ \Theta(\log^* h)$ pebbles in the reversible pebble game, where $\log^*$ is the iterated logarithm function \cite{Kra01}. 
While $\log^*$ is a slow growing function, this gives a provable separation between reversible and irreversible pebbling.
In \cite{KSS15} the authors give a reversible pebbling of the binary tree using $(1+\epsilon)h$ pebbles and $n^{O(\log 1/\epsilon)}$ steps; however, there is no known algorithm for reversibly pebbling the binary tree with the minimum number of pebbles in polynomial time.
We now show that for the binary tree, the bounds for the spooky pebble game are much closer to the irreversible pebble game.
}
{It is known that $h+1$ pebbles and $\Theta(n)$ steps are both necessary and sufficient to pebble a binary tree of height $h$ with $n=2^h-1$ nodes in the irreversible pebble game \cite{Coo73}.
In \cite{Kra01}, Kr{\'a}l'ovi{\v{c}} showed that $(h + \Theta(\log^* h))$ pebbles are necessary and sufficient this task in the reversible pebble game, where $\log^*$ is the iterated logarithm function.
Despite this there are no known polynomial time pebbling strategies for the binary tree using the optimal number of pebbles; the best known tradeoff is a strategy that uses $(1+\epsilon)h$ pebbles and runs in $n^{O(\log 1/\epsilon)}$ steps \cite{KSS15}.
We show the existence of a spooky pebbling strategy for the binary tree that uses the same number of pebbles as the most efficient irreversible pebbling while only being a $O(\log n)$ factor slower.
Thus the spooky pebble game is almost as efficient as the irreversible pebble game and is provably more efficient than the reversible pebble game on binary trees.
}
\begin{theorem}
There exists a spooky pebbling algorithm on the complete binary tree with $n = 2^h - 1$ nodes that uses $h+1$ pebbles and $O(n \log n)$ steps.
\end{theorem}
\begin{proof}
Before getting to the algorithm, we present the following subroutine:
\begin{lstlisting}
fast_pebble$(\calT,h)$: // $\calT$ is the root and h is the height of the tree
    if($h=1$):
        place($\calT$)
    else:
        run fast_pebble($\calT.l,h-1$)
        run fast_pebble($\calT.r,h-1$)
        place($\calT$)
        remove($\calT.l$)
        remove($\calT.r$)
\end{lstlisting}
The above subroutine uses at most $h+1$ pebbles and $2^{h+1}-1$ steps to place a pebble at the root of the tree while leaving ghosts at all other \edit{}{internal} nodes.
We will now use fast\_pebble to construct a recursive unpebbling algorithm for a complete binary tree with root $\calT$ and depth $h$:
\begin{lstlisting}
$\calU(\calT,h):$
    remove($\calT$)
    run fast_pebble($\calT.l,h-1$)
    run fast_pebble($\calT.r,h-1$)
    place($\calT$)
    remove($\calT$)
    remove($\calT.r$)
    run $\calU(\calT.l,h-1$)
    run $\calU(\calT.r,h-1$)
\end{lstlisting}
We note that $\calU(\calT,h)$ uses at most $h+1$ pebbles\edit{ and t}{; we do not need to remove $\calT.l$ before step $8$ as the pebble on this nodes is removed at the beginning of the recursive call. T}he number of steps \edit{it}{this algorithm} takes follows the recurrence relation:
$$T(h) \leq 2T(h-1) + 2^{h+2} +\edit{1}{2}$$
Unrolling the recurrence gives us that $T(h)$ is $O(h2^h)$, which in turn is $O(n \log n)$. We then note that $\calU$ can be converted \edit{in to}{into} a pebbling algorithm $\calP$ by removing step 6. Doing this does not change the number of pebbles used and reduces the total number of steps by one, so $\calP$ also uses at most $h+1$ pebbles and only $O(n \log n)$ steps.
\end{proof}
\edit{Thus the spooky pebble game requires less pebbles than the reversible pebble game for the complete binary tree and comes within a logarithmic factor of the number of steps required for irreversible pebbling with the same number of pebbles.}{}

%% file: future-work.tex
\section{Future work}
\edit{}{
As the main result of this paper, we have shown asymptotically tight upper and lower bounds for the spooky pebble game when played on the line graph.
Going beyond the line graph presents significant challenges as the number of pebbles needed in the spooky pebble game to pebble an arbitrary DAG is PSPACE-hard to approximate.
Despite this, for the binary tree graph with height $h$ and $n=2^h-1$ nodes we were able to construct a time-efficient (i.e. $O(n \log n)$) pebbling strategy that uses the optimal number of pebbles (i.e. $h+1$).
The time optimality of this algorithm is unknown and it would be interesting to construct a more time-efficient algorithm for the binary tree or prove it is impossible.
On the other extreme there is a straightforward spooky pebbling algorithm that uses $n$ pebbles and only $O(n)$ steps.
Is there a general tradeoff between the number of pebbles and the number of steps for the binary tree similar to our results for the line graph?
Beyond binary trees we think it would be interesting to explore strategies for the spooky pebble game on other graph topologies like butterfly graphs used in computing the DFT \cite{Tom80}, and those used in the construction of data-independent memory-hard functions \cite{AS15, RD16, ABP17}.
Finally, the spooky pebble game is but one way to use intermediate measurements to improve the time-space efficiency of quantum algorithms.
Whether other methods of using intermediate measurements could yield better time and space efficiency tradeoffs remains open.
}

%% file: 7-acknowledgements.tex
\section{Acknowledgments}
We would like to thank Craig Gidney for insightful discourse regarding his blog post on the spooky pebble game \edit{. We also want to thank}{as well as} Eddie Schoute and Patrick Rall for discussions about potential applications of the spooky pebble game.
\edit{}{Additionally we would like to thank the anonymous reviewers of our paper for taking the time to review our manuscript. We sincerely appreciate their comments and suggestions, which improved the quality of this manuscript.}